\documentclass[12pt,aas2pp4]{aastex}

\newcommand{\hi}{H\,{\sc i}}
\newcommand{\hii}{H\,{\sc ii}}
\newcommand{\prim}{$^{\prime}$}
\newcommand{\prin}{$^{\prime\prime}$}
\newcommand{\aprox}{${\sim}$}
\newcommand{\Jy}{~Jy~km~s$^{-1}$}
\newcommand{\km}{~km~s$^{-1}$}
\newcommand{\degree}{$^{\circ}$}

\newcommand{\msolar}{$M_{\odot}$}
\newcommand{\lsolar}{$L_{\odot}$}

\newcommand{\mhi}{$M_{\mathrm {HI}}$}

\slugcomment{Resubmitted to AJ}

\shorttitle{\hi\ and CO in Blue Compact Dwarf Galaxies}
\shortauthors{Bravo--Alfaro et al.}

\begin{document}

\title{\ion{H}{1} and CO in Blue Compact Dwarf Galaxies: Haro~2 and Haro~4}

\author{H.\ Bravo--Alfaro}
\affil{Departamento de Astronom\1a, Universidad de Guanajuato, Apdo. Postal
144,  Guanajuato 36000, M\'exico}

\author{E.\ Brinks}
\affil{Instituto Nacional de Astrof\'{\i}sica, Optica y Electr\'onica,
Apdo.\,Postal 51 \& 216, Puebla, Pue 72000, M\'exico }  

\author{A.\ J.\ Baker\altaffilmark{1}}
\affil{Owens Valley Radio Observatory, Caltech 105--24, Pasadena, CA 91125}

\author{F. Walter}
\affil{National Radio Astronomy Observatory, P.O. 
Box O, Socorro, NM 87801}

\author{D. Kunth}
\affil{Institut d'Astrophysique, 98bis Bd. Arago, F--75014 Paris, France}

\altaffiltext{1}{Present address: Max--Planck--Institut f{\" u}r 
extraterrestrische Physik, Postfach 1312, D--85741 Garching, Germany}

\begin{abstract}

We present VLA \hi\ imaging of two blue compact dwarf (BCD) galaxies, Haro\,2
and Haro\,4, and of the spiral galaxy Haro\,26, which is projected some
22\prim\ SW of Haro\,4.  We also show a map of the CO(1--0) distribution of
Haro\,2 obtained with the OVRO millimeter array, as well as derive an
upper limit for CO(2--1) emission from Haro\,4 obtained with the CSO.
The \hi\ data of Haro\,2 
reveal that the kinematical major axis lies perpendicular to the photometric
major axis, indicating that the atomic hydrogen rotates about the major axis
of the galaxy. This confirms earlier indications based on CCD photometry that
Haro\,2 is a dust--lane dE rather than a dIrr. We propose that the present
neutral and molecular ISM configuration is due to recent gas accretion or a
merger. The \hi\ distribution and dynamics of Haro\,4 and the neighboring
spiral Haro\,26 suggest that they are currently undergoing a tidal
interaction, reinforcing the notion that interactions play an important role
in triggering the star formation witnessed in Blue Compact Galaxies.

\end{abstract}


\section{Introduction} \label{s-intro}

Blue compact dwarf (BCD) galaxies are dwarf irregular systems that are
currently experiencing a burst of star formation.  Their appearance is
dominated by the starburst, which appears compact and blue, hence their name.
Their spectra show all the characteristics of \hii\ regions, which has led
many BCDs to be classified as \hii\ galaxies (although, in general, not all
\hii\ galaxies are necessary dwarf systems).  At their current star
formation rates (SFRs), these systems would exhaust their total available gas
in considerably less than a Hubble time; for this reason, it is generally
accepted that their star formation must be episodic.

Gas--rich dwarf galaxies play an important role in our attempts to understand
galaxy formation and evolution.  BCDs show low metallicities which
suggests that they are chemically unevolved systems.  They are more
metal--deficient than the solar neighborhood by factors between three and
twenty \citep{kunt00}, with a few objects such as I\,Zw\,18 \citep{sear72}
and SBS\,0335--052 (Melnick, Heydari--Malayeri, \& Leisy 1992) reaching
metallicities as low as $0.025\,Z_\odot$.

This said, most BCDs do not seem to be young objects in the sense that
they are undergoing their first episode of star formation.  Many reveal older
stellar populations underlying the bright young stars produced in the current
bursts \citep{papa96,oest00,tosi01}.  This poses some interesting questions
regarding the production and possible loss of heavy elements.  Recent X--ray
observations with the {\em Chandra X--ray Observatory} seem to indicate that
in starbursting dwarf galaxies such as NGC\,1569 and NGC\,3077 (Martin,
Kobulnicky, \& Heckman 2002; Ott, Martin, \& Walter 2003), blow--out of
material into the halo occurs that will remove the products of
nucleosynthesis in stars more massive than 8\,\msolar\ and disperse those
across a larger volume. Numerical simulations have suggested that in extreme 
cases blow--away can occur, which removes these products from the galaxy 
proper and enriches the intergalactic medium \citep{macl99}.

It is not clear what causes a BCD to initiate a burst or episode of star
formation, and several mechanisms have been proposed, ranging from internal
instabilities to external (especially tidal) triggers.  Reports that some
BCDs are isolated systems (e.g., van Zee, Skillman, \& Salzer (1998), who
observed four such objects in \hi) suggests that interactions may not be the
only explanation.  But other studies (e.g., Taylor et al. 1994; Taylor 1997;
Walter et al. 1997; Pustilnik et al. 2001b, and references therein) have
shown that the presence of close companions producing tidal interactions may
be a sufficient condition to explain BCD bursts.  Furthermore, recent studies
(see e.g., Noeske et al. 2001) of large samples of star--forming dwarf
galaxies that look for faint companions give additional support to the
hypothesis of interaction--induced star formation in BCDs. 

Most studies to date have focused, broadly speaking, on the optical
characteristics of BCDs, with several incorporating \hi\ observations of
sufficiently high spatial resolution to resolve the objects under study.
Despite the fact that gas has to pass through a molecular phase for star
formation to commence, only very few authors have reported high--resolution
(interferometric) maps of molecular gas in BCDs.  Detecting CO in dwarf
galaxies has proved to be quite a challenge, a fact that is frequently
attributed to the low metallicity in these objects (Gondhalekar et al. 1998;
Taylor, Kobulnicky, \& Skillman 1998; Barone et al. 2000; Meier et al. 2001b).
As a result, only a few dwarfs have been clearly detected in CO, and still
fewer objects have been mapped with millimeter interferometers (NGC\,5253: 
Turner, Beck, \& Hurt 1997; Meier, Turner, \& Beck 2002; NGC\,1569: Taylor et 
al. 1999; NGC\,4214: Walter et al. 2001; NGC\,3077: Meier, Turner, \& Beck 
2001a; Walter et al. 2002; IC\,10: Walter et al. 2004. in preparation).
With the advent of the  
FUSE telescope, $\rm H_2$ molecules have been searched for in BCDs, but upper 
limits for the diffuse $\rm H_2$ column density are found to be very low, with 
$N(\rm H_2$)$\lesssim10^{15}$\,cm$^{-2}$ \citep{vida00}.

To increase the sample of galaxies studied in reasonable detail, we have
mapped the BCDs Haro\,2 (= Mkn\,33 = UGC\,5720) and Haro\,4 (= Mkn\,36) in
\hi. In addition, we have obtained an interferometric map of Haro\,2 in the
CO(1--0) line.  Haro\,2 is catalogued as a Wolf--Rayet galaxy (Beck, Turner,
\& Kovo 2000; Kunth \& Joubert 1985) and has been extensively studied as a
key object for understanding star formation activity in low--metallicity
environments.  It is also one of the few BCDs that has been detected in
CO(1--0), CO(2--1) and CO(3--2) single--dish observations (Sage et al. 1992;
Israel, Tacconi, \& Baas 1995; Barone et al. 2000; Meier et al. 2001b).

This paper is organized in the usual way.  In Section \ref{s-obs}, we describe 
the \hi\ and CO observations.  The \hi\ distributions and kinematics of the 
galaxies are shown in Section \ref{s-res}, where we also present the results 
derived from the CO observations.  In Section \ref{s-disc} we discuss the 
atomic and molecular gas contents and distributions of Haro\,2 and Haro\,4,
and the implications for BCD evolution.  A summary of our results is
presented in Section \ref{s-summ}.

\section{Observations} \label{s-obs}

\medskip
\centerline{EDITOR: PLACE TABLE \ref{tab-optparm} HERE.}
\medskip

Table~\ref{tab-optparm} summarizes the optical properties of the two galaxies
under study plus some relevant data on Haro\,26 (= NGC\,3510), which lies at
a separation of \aprox22\prim\ southwest of Haro\,4.  We mapped the \hi\ in
Haro\,2 and the Haro\,4/Haro\,26 system with the NRAO\footnote{The National
Radio Astronomy Observatory is a facility of the National Science Foundation,
operated under cooperative agreement by Associated Universities, Inc.} Very
Large Array (VLA). Haro\,2 was mapped in the CO(1--0) rotational transition
with the Owens Valley Radio Observatory (OVRO) millimeter array.
Single--dish CO(2--1) observations of Haro\,4 were obtained at the Caltech
Submillimeter Observatory (CSO).  In this section we describe the procedures
used to obtain and reduce the data.

\subsection{\ion{H}{1} imaging} \label{ss-hiobs}

The \hi\ data were obtained with the VLA in its C and D configurations,
during two runs carried out in November 1994 and in May 1995 respectively
(see Table~\ref{tab-loghi} for the VLA observing log). Two pointings were
used for each run, one centered on Haro\,2 and the other on Haro\,4, but
including within the primary beam the galaxy Haro\,26.  We observed with the
correlator in mode 4, resulting in two datasets---one with 63 channels at
5.2\,\km\ resolution, and the other with 31 channels at 20.8\,\km\
resolution.  Hanning smoothing was applied on--line during the observations.
After inspecting both datasets, we decided to work only with the lower
velocity resolution data because of insufficient signal--to--noise at high
velocity resolution.  Both C-- and D--array data were combined, leading to a
spatial resolution of about 14\prin$\times$15\prin.

\medskip
\centerline{EDITOR: PLACE TABLE \ref{tab-loghi} HERE.}
\medskip

Standard VLA calibration and imaging procedures were applied, using the NRAO
astronomical image processing system (AIPS, see van~Moorsel, Kemball \&
Greisen (1996) for an up--to--date description).  Data cubes were produced
with robust weighting \citep{brig95} in order to maximize sensitivity while
preserving as high a spatial resolution as possible. In order to define the
extent of the \hi\ emission, we initially averaged four line--free channels
on either side of the band to create a first approximation of the continuum.
The corresponding image was subtracted from the channel maps, resulting in a
cube of \hi\ line emission only.  We used this cube to conduct a more
thorough search for 21~cm line emission.  Following this search, a new
continuum image was made using all channels deemed free from line
emission.  After continuum 
subtraction, the channels containing line emission were CLEANed.  In order to
build moment maps from the data cubes we used a masking technique: for each
CLEANed cube with a beam of around 15\prin, we applied a convolution to a
beam approximately twice as large, i.e., 30\prin.  The smoothed cubes were
then clipped at a level of $2\sigma$, and \hi\ emission was identified via
visual inspection on the basis of spatial continuity in neighboring channel
maps.  A conditional transfer (or {\it mask}) was applied to the original
CLEANed cube using the blanked, smoothed data.  As a last step \hi\
surface brightness maps and velocity fields were created.

\medskip
\centerline{EDITOR: PLACE TABLE \ref{tab-obshi} HERE.}
\medskip

The final cubes for Haro\,2, targeted at the center of the observed field,
have an rms per channel of $0.37\,{\rm mJy\,beam^{-1}}$. The cubes for Haro\,4
and Haro\,26 have an rms per channel of $0.39\,{\rm mJy\,beam^{-1}}$ at the
pointing center, which lies close to the position of Haro\,4.  At the 
position of Haro\,26, the correction for primary beam attenuation 
elevates the rms by a factor of  5, approximately.  The
observations have on average an \hi\ mass detection threshold for individual
clouds of $4.5 \times 10^{6}\,M_\odot$ and $1.1 \times 10^{6}\,M_{\odot}$ for
Haro\,2 and Haro\,4 respectively.  These values correspond to surface
brightness sensitivities of $2.5 \times 10^{20}\,{\rm cm}^{-2}$, where we
have assumed that a signal is real if it shows up at the $3\sigma$ level over
at least two consecutive channels. As a byproduct of the \hi\ observations,
we obtain useful information about the radio continuum emission coming from
the three galaxies as well (see below). The \hi\ observational parameters of
the combined C-- and D--array data are listed in Table~\ref{tab-obshi}, where
columns~2 and 3 give the beam size in seconds of arc and the linear
resolution in kpc, respectively. Column~4 gives the channel width, column~5
the rms noise per channel after continuum subtraction, and column~6 the
noise expressed in units of Kelvin brightness temperature.

\subsection{CO observations} \label{ss-coobs}

We observed the CO(1--0) transition in Haro\,2 with the L and H
configurations of the OVRO millimeter array \citep{padi91} during the
1995--96 season.  Table~\ref{tab-logco} summarizes our observations.  The
array includes six 10.4--m diameter antennas with half--power beamwidths of
$60\arcsec$ at 115\,GHz.  Two separate correlators processed continuum and
spectral line data: an analog correlator recorded continuum bandwidths of
1\,GHz centered at $\pm 1.5\,\rm{GHz}$ relative to the reference local
oscillator frequency, while a digital correlator provided 112 frequency
channels in the upper sideband, Hanning--smoothed to 4\,MHz
($10.4\,\rm{km\,s^{-1}}$) resolution.

\medskip
\centerline{EDITOR: PLACE TABLE \ref{tab-logco} HERE.}
\medskip

We calibrated the data within the OVRO millimeter array database using the
MMA package \citep{scov93}.  Observations of Uranus obtained with similar
elevation and atmospheric decorrelation were used to determine the flux
density of the bright quasar 1150+497 at each epoch; we estimate that the
resulting uncertainty in our flux scale is $\sim10\%$.  Paired observations
of 1150+497 interleaved with observations of Haro\,2 every 30--40 minutes
were then used to remove the effects of instrumental and atmospheric
variations on phase and amplitude gains.  Observations of bright quasars were
used for passband calibration.

After editing the visibilities in the DIFMAP package \citep{shep97}, our 
final dataset included 24.3 hours of integration and 30 distinct baselines.  
We mapped the data using the IMAGR task in AIPS, cleaning a single large 
region down to the rms noise in each velocity channel.  Maps of the 1\,GHz 
continuum channel in the line--free sideband showed no evidence of emission 
(with a $3\sigma$ limit of 2.3\,mJy at 112\,GHz), so subtraction of continuum 
from the spectral line data was unnecessary.  Natural weighting produced a 
synthesized beam of $3.3\arcsec \times 2.6\arcsec$ at position angle 
$-84.0^\circ$, and an rms noise per channel of $9.6\,{\rm mJy\,beam^{-1}}$.

We observed Haro\,4 in the CO(2--1) transition at the Caltech Submillimeter
Observatory (CSO) on Mauna Kea, Hawaii on 2001 April 8.  The beam size of the
CSO at 230 GHz is \aprox30\prin, and the main beam efficiency is determined
from observations of Mars to be $\eta_{\rm mb}=0.72$ \citep{geri00}. The 
1024--channel 500\,MHz AOS spectrometer was used for the observations.  The 
system temperature was 630~K ($\tau_{225\,{\rm GHz}} = 0.2$).  In total, 
1800\,s were spent on--source, leading to an rms of $T^*_A = 9\,{\rm mK}$ 
($T_{\rm mb} = 12.5\,{\rm mK}$) in a $6\,{\rm km\,s^{-1}}$ Hanning--smoothed 
channel.  No signal was detected in our observations.

\section{Results} \label{s-res}

\subsection{\ion{H}{1} morphology and kinematics of Haro\,2} \label{ss-h1inh2}

This is the first extensive interferometric \hi\ work on Haro\,2; previous
\hi\ results were restricted to single--dish studies.  Like most
star--forming dwarf galaxies, Haro\,2 is \hi--rich. Figure~1
presents the channel maps.  These show clearly resolved structures that are
individually oriented in a southeast--northwest direction, and that show a
slight displacement perpendicular to this direction as a function of
velocity.

The \hi\ parameters we obtain are listed in Table~\ref{tab-hiparm}.  Column~1
gives the galaxy name; columns~2 and 3 list the velocity centroid and 
FWHM.  We conservatively take half the channel separation (10.3\,\km) as the
uncertainty for the \hi\ central (systemic) velocity.  The error in the
velocity width is estimated to be of order a quarter of a channel, or
$\approx 5$\,\km.  The integrated \hi\ flux (corrected for primary beam
attenuation) is listed in column~4; we estimate an error of about 10\% for
the absolute flux calibration.  The radio continuum flux determined from the
line--free channels is given in column~5, and the integrated \hi\ mass in
column~6. Column~7 gives the observed peak \hi\ column density.

\medskip
\centerline{EDITOR: PLACE TABLE \ref{tab-hiparm} HERE.}
\medskip

Comparison of our \hi\ detection with published single--dish data yields the
following results.  \citet{gord81} observed Haro\,2 with the 300--ft Green
Bank Telescope, which had a 10.8\prim\ beam.  From examining their
published spectra, we deduce a peak flux of 34\,mJy and a line width at 50\%
of the peak of $\Delta v_{50} \approx 100$\,\km. The authors publish a line
width at 20\% of the peak of $\Delta v_{20} = 208$\,\km\ and a total \hi\
mass of 5.2$\times$10$^8$\,\msolar\ (using a distance of 21\,Mpc).  If we
adjust their distance to the one used throughout this paper (19.5\,Mpc), the
total \hi\ mass is $4.2 \times 10^8$\,\msolar. In a more recent study with
the Effelsberg 100--m telescope, which gives a 9.3\prim\ beam, Huchtmeier,
Sage, \& Henkel (1995) report ${\Delta} v_{50} = 110\,{\rm km\,s^{-1}}$,
${\Delta} v_{20} = 170\,{\rm km\,s^{-1}}$, and a peak flux of $34.0\pm
3.4$\,mJy. These authors point out that their observations of Haro\,2 may
suffer from confusion with a close companion located 4.7\prim\ to the east
(see also Nilson 1973).  We inspected our data cube around the position of
the putative companion but detected no \hi\ emission.  It is therefore
justified to use their single--dish data to derive an \hi\ mass for Haro\,2,
which (integrating the published spectra by hand) comes out to be $3.6 \times
10^8$\,\msolar.

The two single--dish results agree with each other within their respective
uncertainties.  The single--dish linewidths also agree with the value we
determine from our interferometric data, but the latter fall short as far as
the integrated flux is concerned: the VLA recovers an \hi\ mass of order $2.3
\times 10^8$\,\msolar, i.e., only $\sim 55\%$ of the single--dish mass (see
Table \ref{tab-hiparm}).  We are aware of the fact that CLEANed maps do not
always yield accurate integrated fluxes, especially when the \hi\
distribution in the channel maps is extended on scales comparable to or
larger than $\sim 50\%$ of the primary beam (see, e.g., J\"ors\"ater \& van
Moorsel 1995).  In the case of Haro\,2, whose \hi\ emission has an extent of
order $1^\prime - 2^\prime$, or less than 1/15th of the primary beam, we
discard this as a possible explanation.  Instead, we postulate that the VLA
maps only reveal the ``tip of the iceberg'' and that the galaxy contains
extended, low surface brightness gas that is mostly missed by the
interferometer.  Figure~2 shows the integrated spectrum from our
VLA data.  This profile is in very good agreement with that obtained by
\citet{huch95} in terms of both the velocity width and the slight asymmetry
towards lower velocity side, where a bump appears near 1420\,\km.  We do not
confirm the secondary peak seen by \citet{gord81} at 1550\,\km; this feature
was also not detected by \citet{huch95} at Effelsberg. In
Figure~2 it is also clear that we are missing part of the
extended emission of the galaxy.

Figure~3 is a map at $15^{\prime\prime}$ resolution of the \hi\
surface brightness in Haro\,2 superposed on an optical DSS image in grey
scale. It appears elongated in a southeast--northwest direction, much like
what can be seen in the channel maps. The \hi\ disk dimensions are
$1.4^\prime \times 0.7^\prime$ (equivalent to $7.9\,{\rm kpc} \times
4.0\,{\rm kpc}$ at the distance of Haro\,2), at a level of $\sim 5 \times
10^{20}$\,cm$^{-2}$. These are estimated along the optical major and minor
axis of the galaxy, respectively, as shown by the two lines drawn in
Figure~3, but were measured from a smoothed \hi\ map that is not
shown in this work. The \hi\ disk orientation roughly coincides with the
$1.0^\prime \times 0.9^\prime$ ($5.7\,{\rm kpc} \times 5.1\,{\rm kpc}$)
optical disk, whose major axis has a position angle of approximately
135\degree. At least three features appear to be unresolved \hi\ clouds with
optical counterparts: one to the west of Haro\,2, (labeled A, at
$\alpha_{2000}\,=\,10^{h}32^{m}27^{s}$,
$\delta_{2000}$\,=\,+54\degree24\prim00\prin), one to the east (labeled B, at
$\alpha_{2000}\,=\,10^{h}32^{m}37^{s}$,
$\delta_{2000}$\,=\,+54\degree24\prim00\prin), and one to the northeast
(labeled C, at $\alpha_{2000}\,=\,10^{h}32^{m}37^{s}$,
$\delta_{2000}$\,=\,+54\degree24\prim43\prin).  These \hi\ concentrations may
either be dynamically distinct structures, or represent only the local peaks
of a smooth extended \hi\ component that the VLA otherwise resolves out.  In
either case, although each of the peaks is detected in at least two velocity
channels, more sensitive observations will be needed to confirm their
reality.

Figure~4 presents the \hi\ velocity field of Haro\,2 and reveals
the galaxy's complex kinematics.  There is virtually no velocity gradient
visible along the major axis defined by the optical light and \hi\
distribution; a cursory inspection of the channel maps (Figure~1) 
already hints at this behavior.  A clear gradient is seen at roughly right 
angles to the major axis, with only a weak suggestion of rotation about the 
minor axis to the east.  The low velocity resolution of our observations 
unfortunately prevent us from confirming this latter feature.  In 
general, the kinematics of Haro\,2 inferred from its \hi\ emission is 
consistent with the analysis of the inner $30\arcsec$ by \citet{legr97} using 
the H$\alpha$ emission line.  

\subsection{The molecular gas in Haro\,2} \label{ss-h2inh2}

Figure~5 shows the CO(1--0) total intensity map for Haro\,2,
integrated over twelve channels from 1378 to $1503\,{\rm km\,s^{-1}}$ LSR.
We detect a clear arclike feature extending from southeast to northwest,
containing several distinct emission peaks.  Taken together, these cover the 
same area on the sky as that in which vigorous star formation is seen in the 
optical, as the overlay in Figure~6 shows.  There appears to be an
additional concentration of gas $\sim 10^{\prime\prime}$ to the south, and a
possible component just as far to the northwest (labeled S and NW in 
Figure~5); these clouds are also seen in independent data obtained 
with the IRAM Plateau de Bure interferometer by \citet{frit00}.

After clipping the data cube at the $1\sigma$ level, we corrected for the
primary beam response and measured the galaxy's total CO(1--0) flux to be
$F_{\rm CO} = 24.3 \pm 3.2\,{\rm Jy\,km\,s^{-1}}$.  Excluding the outlying
concentrations of gas, we find $F_{\rm CO} = 17.6 \pm 1.7\,{\rm
Jy\,km\,s^{-1}}$.  Assuming a Galactic CO--to--${\rm H_2}$ conversion factor
of $X_{\rm CO} \equiv N_{\rm H_2}/I_{\rm CO} = 2 \times 10^{20}\,{\rm
cm^{-2}\,(K\,km\,s^{-1})^{-1}}$ \citep{stro88,hunt97}, we estimate the total
molecular hydrogen masses following Sanders, Scoville, \& Soifer (1991):
{\small
\begin{equation} \label{e-mh2}
M_{\rm H_2} = 1.18 \times 10^4\,M_\odot\,\Big({\frac {X_{\rm CO}} {3 \times 
10^{20}}}\Big)\,\Big({\frac D {\rm Mpc}}\Big)^2\,\Big({\frac {F_{\rm CO(1-0)}}
{\rm Jy\,km\,s^{-1}}}\Big)
\end{equation}
}
The observed line fluxes then imply $M_{\rm H_2} = (7.3 \pm 0.9) \times
10^7\,M_\odot$ and $(5.3 \pm 0.5) \times 10^7\,M_\odot$ for all of Haro\,2
and for its central regions, respectively.  Multiplying by an additional
factor of 1.36 to account for helium will give the total masses of gas
associated with the CO--emitting structures.  Use of a metallicity--dependent
$X_{\rm CO}$ (Wilson 1995; Arimoto, Sofue, \& Tsujimoto 1996) would elevate
these gas masses by an additional factor of 3--4 for Haro\,2's $Z \simeq
0.3\,Z_\odot$.  However, more recent studies of M\,33 \citep{roso03} and
metal--poor dwarf galaxies (e.g., Walter et al. 2004, in preparation) suggest
that $X_{\rm CO}$ does not depend on metallicity.  Moreover, the results of
the study of molecular gas in Haro\,2 by \citet{frit00} validates the use of
a nearly Galactic conversion factor to estimate $M_{\rm H_2}$ in this
particular system.

Since interferometric data are insensitive to emission which is smoothly
distributed on scales of order that of the primary beam, we wish to estimate
the fraction of the total line flux our map recovers.  We have therefore
corrected the entire data cube for the primary beam response, convolved it to
the half--power beamwidths of $55\arcsec$ and $22\arcsec$ for the NRAO 12\,m
and IRAM 30\,m telescopes, and measured the integrated intensities of our
convolved data at the pointing centers used for published single--dish
observations.  We find we have recovered $\sim 56\%$ of the the total
CO(1--0) emission seen by \citet{isra95} in their $55\arcsec$ beam, and a
mean (for four separate pointings) of $\sim 61\%$ of the flux seen by
\citet{sage92} and \citet{baro00} in their $22\arcsec$ beams.  We make no
attempt to correct for this undetected flux in what follows.

A comparison between the CO (greyscale) and \hi\ (contours, as in
Figure~3) is shown in Figure~7, revealing interesting
common features.  The lower resolution obtained for the \hi\ relative to the
CO prevents a detailed comparison in the very center of Haro\,2 where the CO
displays a large degree of structure.  In spite of this limitation, we
confirm that the arc--like CO feature follows a distribution along the same
(southeast--northwest) direction as the \hi; both are coincident with the
major axis of the galaxy seen in blue light. The CO is more concentrated,
with a linear size smaller than the stellar component, whereas the \hi\
appears more extended.  With the exception of the southeast tail, the
brightest CO regions are projected onto the innermost \hi\ contour where star
formation is observed to be most active (see also Section \ref{s-disc}).

A more detailed comparison is possible between the CO map and the $\sim
3\arcsec$ resolution {\it ROSAT} HRI X--ray image published by Summers,
Stevens, \& Strickland (2001).  Figure~8 shows an overlay of CO
contours on a greyscale representation of the X--ray emission.  Given that
residual {\it ROSAT} pointing errors (which dominate the uncertainty in the
relative astrometry of these two images) can be as large as $10\arcsec$, the
agreement in the location and spatial extent of the highest surface
brightness structures at both wavelengths (as well as in the optical; see
Figure~6) is quite good.  In particular, there is a tantalizing alignment
between the CO arc and the lower of the two X--ray ridges defining a tilted
``V'' morphology in the central part of the HRI image.  This coincidence can
be explained if part of the X--ray emission is produced by shock--heated gas
at the interface between an expanding superbubble and a collection of dense
molecular clouds.  The comparable spatial extent of ${\rm H\alpha}$ emission
in Haro\,2 \citep{mend00} and evidence for an expanding shell in ionized gas
kinematics \citep{lequ95,legr97} offers further support for this picture.
Alternatively, the X--ray emission could be a consequence of the active star
formation that is taking place currently and the associated detonation of
core--collapse supernovae.  The formation of high mass X--ray binaries was
deemed less likely by \citet{summ01}, based on their hardness ratio as
compared to that of the extended emission seen in Haro\,2.  Higher resolution
X--ray observations with instruments like {\em XMM-Newton} or {\em Chandra}
will be needed to determine conclusively the nature of the X--ray emission
and its relation to the ionized, atomic, and molecular ISM.

Figure~9 shows the channel maps for the CO(1--0) line in
Haro\,2, obtained after smoothing down the data from the original $10.4\,{\rm
km\,s^{-1}}$ resolution by a factor of three.  Although the signal--to--noise
for many of the features is rather meager, it is clear that the kinematics of
the molecular gas is quite complex.  The central arc--like feature in the
integrated intensity map (Figure~5) does not appear to have a
coherent velocity structure.  Moreover, the spatially outlying structures do
not seem to be moving at the higher relative velocities expected in the case
of ordered rotation (e.g., the southeast component appears mainly in the
$1426.6\,{\rm km\,s^{-1}}$ channel).  As on the larger spatial scales traced
by \hi, the gas kinematics in the nucleus of Haro\,2 are not those of a
quiescent system.

\subsection{Haro\,4 and Haro\,26} \label{ss-h4h26}

Figure~10 shows the \hi\ channel maps for Haro\,4.  Despite the fact that it
is almost twice as close as Haro\,2, its \hi\ distribution is only marginally
resolved at our 14\prin~resolution (equivalent to \aprox0.5\,kpc).  The
integrated \hi\ map superposed on an optical DSS image is shown in Figure~11.
This map confirms the highly compact nature of the \hi\ distribution, even
though it is still almost three times more extended than the stellar
component (see Table \ref{tab-ratios}).  No clear rotation pattern is
detected for this galaxy, and higher--resolution VLA observations are
required to understand the kinematics of this object in more detail. The \hi\
content we derive for this galaxy (\mhi$= 0.19 \times 10^8$\,\msolar) agrees
with the single--dish value reported by \citet{gord81}, confirming that no
\hi\ emission is missed by the VLA.

As mentioned in Section \ref{ss-coobs}, no CO(2--1) emission was
detected from Haro\,4.  From the per--channel rms of 9\,mK ($T^*_A$),
the corresponding conversion factor $50\,{\rm Jy\,K^{-1}}$ for 
the CSO at 230\,GHz \citep{geri00}, and an assumed velocity width of
$42\,{\rm km\,s^{-1}}$ (matching the \hi\ FWHM), we derive a $3\sigma$
upper limit on the integrated CO(2--1) flux from Haro\,4 of $\leq
21\,{\rm Jy\,km\,s^{-1}}$.  If the CO(2--1)/CO(1--0) intensity ratio
is $\sim 0.5$ in temperature units, as is the case for the two BCDs
(Haro\,2 and Mrk\,297) mapped by \citet{sage92}, then we would expect
the CO(1--0) line flux from Haro\,4 to be $\leq 43\,{\rm
Jy\,km\,s^{-1}}$.  Equation \ref{e-mh2} then implies that Haro\,4 has
a molecular hydrogen mass $M_{\rm H_2} \leq 1.6 \times 10^7\,M_\odot$.
Correction of $X_{\rm CO}$ for the possible dependence on metallicity
according to the prescriptions of \citet{wils95} and \citet{arim96} could
increase this upper limit on $M_{\rm H_2}$ by factors of 8 and 16,
respectively.

Projected some 22\prim\ SW of Haro\,4, the neighboring spiral galay Haro\,26
is well--resolved by the VLA beam and spans a much larger range in velocity.
The channel maps are presented in Figure~12.  This object is a clear example
of a galaxy in differential rotation seen almost edge--on.  \hi\ emission
extends from roughly 600 to 800\,\km.  Figure~13 shows a composite plot of
integrated \hi\ profiles of Haro\,4 and Haro\,26. The total \hi\ surface
brightness map presented in Figure~14 shows the integrated \hi\ emission of
Haro\,26 superposed on a DSS image. As is common, the \hi\ extends well
beyond the optical image, in this case by about a factor of 1.5 (see Table
\ref{tab-ratios}).  We measure a total \hi\ flux for Haro\,26 of 40\Jy, in
good agreement with previous \hi\ line observations.  \citet{tiff88} observed
this galaxy with the 91--m NRAO telescope and obtained an \hi\ flux of
48.3\,\Jy, very close to the $49.4\,{\rm Jy\,km\,s^{-1}}$ reported by
\citet{hayn99}.  More recently, \citet{tayl94} observed this galaxy with the
VLA and reported an \hi\ mass of $10.1 \times 10^8$\,\msolar\ (scaled to the
our assumed distance of 7.9\,Mpc), only 15\% larger than the value of $8.5
\times 10^8$\,\msolar\ that we report in this work (see Table
\ref{tab-hiparm}).

Figure~15 shows both Haro\,4 and Haro\,26 in a single map at larger scale.
Their projected separation is \aprox45\,kpc. The galaxy seen on the DSS image
midway between Haro\,4 and Haro\,26 is CGCG\,155--052, a background galaxy at
$z \sim 0.03$. The conspicuous \hi\ extension to the NE of Haro\,26,
previously reported by \citet{tayl94}, is contiguous with the rest of the
galaxy in the velocity field (Figure~16).  The galaxy appears symmetrical
only out to a radius of \aprox1\prim\ (some 2.3\,kpc), beyond which the
northern tip veers northeast towards Haro\,4.  The radio continuum map of
Haro\,26 (Figure~17) shows emission coinciding with this northeast \hi\
feature.  To the south, isolated radio continuum blobs extend farther than
the \hi, with intensities at the $\sim 2.5\sigma$ level.

\section{Discussion} \label{s-disc}

\subsection{Haro 2} \label{ss-disc-h2}

Haro\,2 is a peculiar member of the BCD class.  Most BCDs are gas--rich,
dwarf {\it irregular} systems manifesting bursts of star formation.  As first
pointed out by \citet{loos86}, however, Haro\,2 has an $R^{1/4}$ rather than
exponential surface brightness profile in the optical (see also Cair{\' o}s
et al. 2001), leading those authors to classify Haro\,2 as a dwarf {\it
elliptical}.  Haro 2 has a small \ion{H}{1}--to--optical diameter ratio,
$D_{\rm H\,I}/D_{\rm opt} \sim 1.4$, whereas the \hi\ distributions of most
BCDs are usually considerably larger than their optical disks ($D_{\rm
H\,I}/D_{\rm opt} \sim 2 - 5$: Taylor et al. 1994; van~Zee et al. 1998;
van~Zee, Salzer, \& Skillman 2001).  The \hi\ content of Haro\,2 is normal
relative to other BCDs, in contrast to its $M_{\rm H\,I}/L_B$ ratio (see
Table~\ref{tab-ratios}), which is at the low end for BCDs
\citep{smok00,lee02,salz02}.  Of the sample studied by \citet{vanz98,vanz01},
Haro\,2 is most similar to UM\,38 and Mrk\,324, two other dwarf ellipticals
\citep{doub97,cair01} with comparable atomic gas properties.  UM\,465
resembles Haro\,2 in that it is a dwarf elliptical detected in both CO and
\hi\ emission, although it has a somewhat smaller \hi\ mass and somewhat
larger metallicity \citep{sage92}.

While Haro\,2 does resemble a BCD in terms of its 
optical morphology and gas content, it is almost unique in the kinematics of 
its neutral atomic and molecular gas.  Whereas in disk--dominated systems the 
kinematic and isophotal major axes are aligned, in the case of Haro\,2 they 
are nearly perpendicular.  One of the few dwarf galaxies with equally 
pathological kinematics is NGC\,5253, whose own molecular and atomic gas 
components also appear to be predominantly rotating about the galaxy's major 
axis \citep{kobu95,meie02}.  This configuration has been postulated to result 
from a tidal interaction, or from the accretion of gas from a formerly 
gas--rich (now stripped) companion.  Its kinematic complexity, together with 
the presence of a dust lane (like that seen in Haro\,2: M{\" o}llenhoff, 
Hummel, \& Bender 1992), suggest NGC\,5253 may be a dwarf version of a giant, 
post--interaction, dust lane elliptical \citep{oost02}.  In the context of 
this interaction scenario, some of the similarities and differences between 
Haro\,2 and NGC\,5253 deserve special attention.

In terms of star formation, NGC\,5253 appears to be undergoing a burst that 
is markedly younger and more intense than that in Haro\,2.  While the youngest 
stars in Haro\,2 have ages $\sim 10\,{\rm Myr}$ (Fanelli, O'Connell, \& Thuan 
1988), population synthesis fits to star clusters in NGC\,5253 suggest ages in 
the 1--8\,Myr range \citep{trem01}, and its extremely luminous and deeply 
embedded ``supernebula'' (Turner, Beck, \& Ho 2000; Gorjian, Turner, \& Beck 
2001; Turner et al. 2003) must be younger still.  Haro\,2's older burst age 
is also indicated by its lower millimeter/centimeter continuum flux ratio.  In 
particular, our $3\sigma$ upper limit of 2.3\,mJy on 112\,GHz emission from 
Haro\,2 (see Section \ref{ss-coobs} above) implies that no more than 
$3.5\,{\rm mJy}$ of its 24.6\,mJy observed at 1.4\,GHz can be due to optically 
thin free--free emission (assuming a spectral index $\alpha = -0.1$).
Virtually all of  
the 1.4\,GHz flux density from NGC\,5253, on the other hand, is thermal 
(Turner, Ho, \& Beck 1998; Meier et al. 2002).  Finally, the star formation 
efficiency in NGC\,5253 is considerably higher, with its $L_{\rm IR}/M_{\rm 
H_2} \simeq 800\,L_\odot/M_\odot$ \citep{turn97} exceeding the more modest 
$\sim 50\,L_\odot/M_\odot$ we infer for Haro\,2 (see Section \ref{ss-sf} 
below).

It is tempting to associate the older, less vigorous starburst in Haro\,2 
with a dynamical disturbance that occurred longer in the past than the one 
currently afflicting NGC\,5253.  Several lines of evidence indeed favor this 
interpretation.  First, much of the molecular gas in the vicinity of NGC\,5253 
appears still to be falling inward and is not associated with the regions of 
most intense star formation \citep{meie02}.  Haro\,2, in contrast, has a 
\mbox{CO(1--0)} distribution whose spatial extent broadly matches
that of its star  
formation as traced by ${\rm H\alpha}$, X--ray, and optical continuum 
emission; this agreement suggests that Haro\,2's molecular component has 
finished coalescing towards the bottom of its potential well.  Second, the 
disagreement between the position angles of the \hi\ and optical disks in 
NGC\,5253-- in fact, they are nearly perpendicular on the sky-- represents 
a level of disturbance not seen in Haro\,2, where the \hi\ follows the optical
disk quite well.  Third, we recall from Section \ref{ss-hiobs} that our VLA 
data failed to recover about half of the \hi\ emission detected in 
single--dish observations of Haro\,2 (see Figure~2).  Diffuse, 
extended emission could very well surround the galaxy on $\sim 10\arcmin$ 
scales, having accumulated there in the same way that late infall in more 
massive post--interaction systems can build regular \hi\ disks \citep{oost02}.

Considering all of the above evidence, we conclude that Haro\,2 is most 
likely a somewhat older cousin of NGC\,5253--- a dwarf elliptical galaxy that 
has recently captured gas in a close interaction or merger with an initially 
gas--rich companion.  For Haro\,2, an obvious candidate donor is the putative 
companion lying 4.7\prim\ to the east \citep{huch95}, whose non--detection 
in \hi\ by those authors could be explained if it lost all its gas in the 
interaction.  If at the same redshift, its projected separation from Haro\,2 
would be a mere 27\,kpc; however, there is no redshift estimate for this 
object, and the real distance which separates it from Haro\,2 remains 
uncertain.  Alternatively, \citet{pust01b} propose the galaxy UGC\,5676 as the 
trigger of the burst of star formation.  This object lies much further away 
in projection, some 36\prim\ ($\sim 200\,{\rm kpc}$) northeast of Haro\,2, but 
is known to share roughly the same redshift.  Both neighbors are fainter than 
Haro\,2 (by 3 and 1.6 magnitudes, respectively) and presumably less massive 
as well, consistent with our assumption that one of them would have lost its 
gas mass to Haro\,2 rather than the other way around.  In view of the 
mass--metallicity relation for dwarf galaxies (e.g., Bender, Burstein, \& 
Faber 1992), Haro\,2 should have received an infusion of less enriched gas 
from its less massive companion.  This would tend to have left Haro\,2 with a 
lower metallicity in its interstellar medium than in its stars, although 
exactly how much lower is difficult to predict without knowing what fraction 
of its gas mass originated elsewhere.  We note that the possibility of tidal 
interaction with more than one object at different times should also not be
ruled out.

In the ``passing encounter'' scenario depicted above--- and even if Haro\,2 is 
actually in the final stages of a merger with a gas--rich dwarf--- gas would 
have been captured at a random angle with respect to the major axis of 
Haro\,2, and would have since settled into a relatively stable orbit 
encircling the galaxy's long axis.  The fraction of gas settling further into 
the gravitational potential of Haro\,2 would accumulate and reach densities 
high enough for molecules to form.  Molecular gas spiralling inward would 
naturally shock, leading to the current burst of star formation we are 
witnessing.  

Even if the \hi\ gas is of relatively recent acquisition, its motion is by now 
dominated by the gravitational potential of Haro\,2, any plausible donor  
having passed perigalacticon and no longer exerting any tidal pull.  If we 
assume that the object 27\,kpc to the east was the source of the accreted gas, 
and if we take as an order of magnitude estimate a relative velocity between 
the two objects of 150\,\km\ (a reasonable upper limit, according to Taylor et 
al. 1994), $\sim 180\,{\rm Myr}$ have elapsed since closest approach, i.e., a 
few revolutions of the captured gas in the potential of Haro\,2.  It is 
therefore valid, as a first approximation, to assume for dynamical mass 
estimates that the gas is moving in roughly circular orbits.  We recall that 
the \hi\ and CO velocity fields show that any rotation is within a plane  
more or less perpendicular to the optical major axis.  Given that no value 
for the inclination is known, we can only estimate a lower limit to the 
dynamical mass. The extent of the gas along the kinematical major axis is 
\aprox1.0\prim. Taking a distance to Haro\,2 of 19.5\,Mpc and a projected 
rotational velocity of 62.5\,\km, we find a value of $M_{\rm dyn} =
2.6 \times 10^9$\,\msolar\ and an $M_{\rm dyn}$\,/\,$L_B$ ratio $=
1.0$. Even when taking into account the fact that the dynamical mass
estimate is a lower limit, for any reasonable $M/L$ ratio for the
stars in Haro\,2 there doesn't seem to be a need to invoke any dark
matter. This is contrary to what is generally seen in BCDs which, if
anything, are more Dark Matter dominated than large, spiral galaxies.

Our estimate of 180\,Myr as the time of closest approach leaves ample
time for the \hi\ to turn into $\rm H_2$. Following the same arguments
as given in \citet{brai01}, the time to transform 20\% of atomic
hydrogen into $\rm H_2$ is $t_{20\%} \approx
\frac{10^7}{n_\mathrm{HI}}$ years where $n_\mathrm{HI}$ is the \hi\
volume density \citep[and references therein]{holl71}. A rough
estimate of the latter can be made by taking the average of the \hi\
column density across the area where CO has been detected and assuming
a line--of--sight thickness for the \hi\ of order 200\,pc. The \hi\
volume density thus 
derived is $n_\mathrm{HI} \approx 3\, \mathrm{cm^{-3}}$,
suggesting that the time needed to transform a substantial amount of
atomic to molecular gas is less than $10^7$\,years, much shorter than
the typical time for solar mass stars to form subsequently ($\approx
30$\,Myr). $t_{20\%}$ is an appropriate indicator for the timescale on
which molecular gas forms in Haro\,2 as most of the gas is still
atomic.  Lower 
metallicities and hence lower dust content will decrease linearly the rate
at which this occurs.

The prominent nuclear star formation in Haro\,2 has been analyzed in the UV 
and optical by \citet{lequ95} using {\em HST}.  These authors find evidence 
for gas flowing away from the central \hii\ region at a velocity of 200\,\km, 
notably strong ${\rm Ly}\,\alpha$ emission with a clear P\,Cygni profile and a 
complex velocity structure (see also Legrand et al. 1997).  The column density 
of neutral hydrogen in this outflowing gas is $N(H) \simeq 7 \times 10^{19}$\,
cm$^{-2}$. This is only a small fraction of the column density measured in the 
VLA map, which is of order $2 \times 10^{21}$\,cm$^{-2}$ (averaged over the 
central pixel at a spatial resolution of $\sim 1.4$\,kpc).  The data are best 
explained by assuming that the central starburst has created an expanding 
supergiant shell measuring $\sim 1.8$\,kpc and expanding at 200\km.  In this 
model, the outflowing gas is predominantly ionized by the central star 
cluster, and the total mass involved in the outflow (if distributed in a 
spherical shell immediately surrounding the \hii\ region) is of order 
10$^{7}$\,\msolar.  Unfortunately, the size of this proposed shell is too close
to the resolution limit of the \hi\ observations for us to be able to detect 
its signature in our data cube.

\medskip
\centerline{EDITOR: PLACE TABLE \ref{tab-ratios} HERE.}
\medskip

\subsection{Haro\,4 and Haro\,26} \label{ss-disc-h4h26}

Haro\,4, shown in Figure~11, appears to be highly centrally concentrated in
both its \hi\ and optical light distributions.  The \hi\ mass of this galaxy
is low compared to that of most BCDs previously observed in \hi\
\citep{vanz98,vanz01,pust01b}; we estimate a lower limit for its dynamical
mass of $M_{\rm dyn} = 5 \times 10^8$\,\msolar.  However, as no clear
indications for rotation are seen, this should only be considered indicative.
Other parameters of Haro\,4 are rather typical among BCDs, such as its
\mhi\,/\,$L_B$ and $D_{\rm H\,I}/D_{\rm opt}$ ratios.  \citet{kunt98} find in
Haro\,4 a broad damped Ly$\alpha$ \ absorption profile centered at the
wavelength corresponding to the redshift of the H$\alpha$ emitting gas. They
show that this occurs chiefly because the \hi\ gas is smooth and virtually
static, as confirmed by the new VLA observations, with respect to the ionized
region where Ly$\alpha$ photons originate.

For Haro\,26, the \hi\ map in Figure~14 confirms the strong warping towards
the northeast already observed in the optical \citep{sand94} and in \hi\
\citep{tayl94}.  While the former authors find no nearby companion to explain
the warp, \citet{tayl94} suggest that the emission extending towards the
north--east may in fact be a separate object.  We find no evidence to support
this scenario, however, as our velocity field in the northeast extension
appears to be continuous with that of the main galaxy.  From the channel maps
(Figure~12) and the zeroth and first moment maps (Figures~14 and 16), we
speculate instead that a long--range interaction with Haro\,4 is responsible
for the warp, with Haro\,4 pulling the near side of the disk of Haro\,26
towards it.  A similar case of a tidal interaction between the BCD
SBS\,0335--052 and the galaxy NGC\,1376 was proposed by \citet{pust01a} to
explain the burst of star formation in the former. The projected separation
between the latter two galaxies is about 150\,kpc, almost three times the
projected distance between Haro\,4 and Haro\,26.  We note that the radio
continuum emission of Haro\,26 shown in Figure~17 follows the same warp seen
in \hi.  The intensities of the radio continuum blobs in the south and the
northeast warp reach $0.25\,{\rm mJy\,beam^{-1}}$.

\subsection{Star formation} \label{ss-sf}

Star formation is thought to occur whenever the local gas density exceeds a
certain threshold. This can be an empirical level, as proposed by
\citet{skil87solo}, who found that star formation in dwarf irregulars occurs
for densities above $\sim 5 \times 10^{20}$\,cm$^{-2}$. Alternatively, the
local gas density can cause clouds to become gravitationally unstable
\citep{toom64}. \citet{tayl94} found in their sample that this threshold was
equivalent to a column density of \hi\ of 10$^{21}$\,cm$^{-2}$. Our spatial
resolution is insufficient to independently calculate the threshold value
for Haro\,2 or Haro\,4. Qualitatively, star formation activity is seen in our
maps above  $10^{21}$\,cm$^{-2}$, and the regions of highest \hi\
column density coincide with the galaxies seen in the optical.  It is worth
noting that the CO in Haro\,2 closely follows the highest \hi\ concentration,
indicating that star formation is likely restricted to the very central
regions as reported by \citet{tayl94}.

The radio continuum also traces the recent star formation activity. In the
case of Haro\,2, the radio continuum distribution (obtained as a byproduct of
the 21\,cm spectral line observations) was only marginally resolved and 
rather featureless. Our 20\,cm radio continuum flux (see Table 
\ref{tab-hiparm}) compares reasonably well with the 20\,mJy reported by 
\citet{beck00}, who show a much higher resolution map than ours (VLA A 
configuration) restricted to a zone smaller than the optical galaxy.

Although Haro\,2 is metal--poor, its \hii\ region metallicity at $\sim
0.3\,Z_\odot$ is not extremely low.  This implies that we are not observing its
first burst of star formation and supports the suggestion that whereas the
youngest burst began less than 10 Myr ago, it was preceded by other bursts
\citep{fane88}.  We can estimate the current rate of formation of
massive ($M \geq 5\,M_\odot$) stars using the relations of
\citet{cond92}. In fact, we can use three independent SFR indicators
and check if they are mutually consistent. 

Based on the 1.4\,GHz continuum luminosity, and assuming the radio continuum
is of non--thermal origin (as argued in Section \ref{ss-disc-h2}), we obtain 
from Condon's equation (21)
\begin{equation} \label{e-sfrc}
{\frac {{\rm SFR}_{1.4}\,(M \geq 5\,M_\odot)} {M_{\odot}\,{\rm yr}^{-1}}} =
{\frac {L_{1.4}}{4.0 \times 10^{21}\,{\rm W\,Hz^{-1}}}} 
\end{equation}
yielding $\sim 0.28\,M_\odot\,{\rm yr}^{-1}$ for Haro\,2 if we use $S_{\rm
1.4\,GHz}$ from Table \ref{tab-hiparm}.  Using the ${\rm H\alpha}$
luminosity, Condon's equation (22) states
\begin{equation} \label{e-sfha}
{\frac {{\rm SFR}_{\rm H\alpha}\,(M \geq 5\,M_\odot)} {M_{\odot}\,{\rm 
yr}^{-1}}} = {\frac {L_{\rm H\alpha}}{4.4 \times 10^{34}\,{\rm W}}}
\end{equation}
yielding $\sim 0.39\,M_\odot\,{\rm yr}^{-1}$ for Haro\,2, if we use the total 
$F_{\rm H\alpha} = 3.8 \times 10^{-12}\,{\rm erg\,cm^{-2}\,s^{-1}}$ implied 
by the narrow--band imaging of \citet{mend00}.  Finally, using the FIR
($40 - 120\,{\rm \mu m}$) dust luminosity as obtained by {\em IRAS},
Condon's equation (26) gives
\begin{equation} \label{e-sffir}
{\frac {{\rm SFR}_{\rm FIR}\,(M \geq 5\,M_\odot)} {M_{\odot}\,{\rm 
yr}^{-1}}} = {\frac {L_{\rm FIR}}{1.1 \times 10^{10}\,L_\odot}}
\end{equation}
yielding $\sim 0.33\,M_\odot\,{\rm yr}^{-1}$ for Haro\,2. Note that we used 
the {\it IRAS} Faint Source Catalog flux densities at 60 and $100\,{\rm \mu 
m}$ to derive $F_{\rm FIR}$.\footnote{We would like to mention  that the {\it
total} SFR in Haro\,2 must also take into account stars with masses $M
< 5\,M_\odot$, the details depending on the  
choice of initial mass function (IMF).  Using the calibration by
\citet{kenn98} for a $0.1-100\,M_\odot$ \citet{salp55} IMF, for example, we 
estimate ${\rm SFR}_{\rm H\alpha} \simeq 1.37\,M_\odot\,{\rm yr}^{-1}$ and 
${\rm SFR}_{\rm IR} \simeq 0.93\,M_\odot\,{\rm yr}^{-1}$.}  The good
agreement between these three estimates suggests that dust does not severely
attenuate Haro\,2's observed ${\rm H\alpha}$ luminosity (at least along our
line of sight).  This favorable geometry, together with the disordered
CO(1--0) velocity field in the inner starbursting region, may explain why
Haro\,2 shows a strong ${\rm Ly\,\alpha}$ emission line with a P~Cygni profile
as reported by \citet{lequ95} and seen in new {\it HST}/STIS data
\citep{mash03}. As we emphasized above, the absence of similar ${\rm
Ly\,\alpha}$ emission from Haro\,4 in the same study may be in part related to
that galaxy's much less disturbed velocity field, which renders ${\rm
Ly\,\alpha}$ photons less able to avoid resonant scattering and escape the
galaxy.

\section{Summary} \label{s-summ}

VLA \hi\ observations of Haro\,2 reveal emission extending over a 7.9\,kpc
$\times$ 4.0\,kpc area, roughly coinciding with the optical galaxy (5.7\,kpc
$\times$ 5.1\,kpc) in extent and orientation. The \hi\ mass we see in our
interferometric data is $2.3 \times 10^8$\,\msolar; a single dish spectrum
suggests than that the object is embedded in a smooth, extended \hi\ halo
containing an additional $1.3 \times 10^8$\,\msolar\ of atomic gas. The
\mhi\,/\,$L_B$ ratio is about 0.1 in solar units. Surprisingly, the
kinematical major axis lies almost perpendicular to the optical (photometric)
major axis. This strongly suggest an external source for the \hi\ gas. The
picture emerging from our study is that of Haro\,2 being a dwarf elliptical
galaxy (as already postulated by \citet{loos86} on the basis of their optical
photometry), which has acquired in an interaction or recent merger a
substantial amount of gas. Assuming that this gas has settled into regular
orbits we derive a dynamical mass of $2.5 \times 10^9$\,\msolar\ and an
$M_{\rm dyn}$\,/\,$L_B$ ratio of 1.0. As we have no independent handle on the
inclination, the estimate of the dynamical mass is a lower limit and could be
larger by a factor of a few. Given the uncertainties involved, this would
imply that there is little need to invoke dark matter in Haro\,2.

High--resolution, interferometric CO(1--0) observations with OVRO broadly
agree with the \hi\ data. The molecular gas coincides with the region of
highest \hi\ column density. The orientation of the CO emission is aligned
with the optical and coincides with published H$\alpha$ data. The kinematics,
although rather disturbed, resembles that of the \hi\ velocity
field. Assuming a Galactic conversion factor, we detect $7.3 \times
10^7$\,\msolar\ of molecular hydrogen.

Haro\,4 is unresolved in \hi. We find an \hi\ mass of $1.9 \times
10^7$\,\msolar. A CO(2--1) observation with the CSO yielded an upper limit to
the mass of molecular hydrogen of less than $1.6 \times
10^7$\,\msolar. Haro\,4 seems to be interacting with the nearby spiral galaxy
Haro\,26 (= NGC~3510), the latter object showing a pronounced warp in the
direction of the BCD.

Our results provide additional proof that interactions with neighboring
galaxies are responsible for the onset of the current episodes of star
formation in BCDs. Although the two galaxies under consideration lie in
relatively low--density environments, we conclude that a major accretion
event or merger can account for the nuclear starburst in Haro\,2 as well as
for its peculiar kinematics, whereas a more transient encounter between
Haro\,4 and the neighboring spiral Haro\,26 could explain both the star
formation in the former and the pronounced warp in the latter.

\acknowledgements

We thank Lesley Summers for providing the {\it ROSAT} HRI image used in
Figure~8, and the NRAO for a generous allocation of observing
time. We are grateful to the anonymous referee for various thoughtful
comments that helped to improve  
the paper.  This research has made use of the NASA/IPAC Extragalactic Database
(NED), which is operated for NASA by the Jet Propulsion Laboratory,
California Insitute of Technology.  We have also used the Digital Sky Survey,
produced at the Space Telescope Science Institute, and the Lyon--Meudon
Extragalactic Database (LEDA) supplied by the LEDA team at the
CRAL--Observatoire de Lyon (France).  HBA thanks the CONACyT of Mexico for
its support through grant 129987--E; EB gratefully acknowledges the award of
CONACyT research grant 27606--E.  The authors thank the scientific and
technical staff at OVRO for making our CO observations possible; the OVRO
millimeter array has been supported in part by NSF grant AST 93--14079 and
the Norris Planetary Origins project.  AJB acknowledges support from an NSF
Graduate Research Fellowship.

\clearpage

\begin{deluxetable}{llcclccccl}
\tablenum{1}
\tablecaption{Optical properties of the galaxies observed in \hi.
\label{tab-optparm}} 
\tablewidth{0pt}
\tablehead{
\colhead{} & 
\colhead{} & 
\colhead{} & 
\colhead{} & 
\colhead{Morph.} & 	
\colhead{}	& 
\colhead{$D$} & 
\colhead{12 +} & 
\colhead{} \\
\colhead{ID} & 
\colhead{Alias} & 
\colhead{$\alpha_{\rm J2000}$} &
\colhead{$\delta_{\rm J2000}$} & 
\colhead{type} &  
\colhead{$M_B$} &  
\colhead{(Mpc)} &  
\colhead{log\,[O/H]} & 
\colhead{Ref.} \\
\colhead{(1)} & 
\colhead{(2)} & 
\colhead{(3)} & 
\colhead{(4)} & 
\colhead{(5)} & 
\colhead{(6)} & 
\colhead{(7)} & 
\colhead{(8)} & 
\colhead{(9)}
}
\startdata
Haro\,2 & Mrk\,33  & 10 32 31.9 & +54 24 03 & HII Im\,pec & $-$18.1 & $19.5^a$ 
& 8.4 & 1 \\
Haro\,4 & Mrk\,36  & 11 04 58.5 & +29 08 22 & BCD         & $-$14.7 & $6.9^b$ 
& 7.8 & 2 \\
Haro\,26& NGC\,3510& 11 03 43.6 & +28 53 06 & SB(s)m      & $-$17.2 & $7.9^b$ 
& --- & --- \\
\enddata
\tablenotetext{a}{Assuming $H_0 = 75\,{\rm km\,s^{-1}\,Mpc^{-1}}$.}
\tablenotetext{b}{From \citet{tull88}.}
\tablerefs{(1) \citet{mash99}\\ (2) \citet{izot99}}
\end{deluxetable}


\begin{deluxetable}{lcclcc}
\tablenum{2}
\tablecaption{VLA \hi\ observing log \label{tab-loghi}}
\tablewidth{0pt}
\tablehead{
\colhead{} & 
\colhead{} & 
\colhead{} & 
\colhead{} & 
\colhead{VLA} & 
\colhead{$t_{\rm int}$} \\
\colhead{Field} & 
\colhead{$\alpha_{\rm J2000}$} & 
\colhead{$\delta_{\rm J2000}$} & 
\colhead{Date} & 
\colhead{config} & 
\colhead{(min)} \\
\colhead{(1)} & 
\colhead{(2)} & 
\colhead{(3)} & 
\colhead{(4)} & 
\colhead{(5)} & 
\colhead{(6)} 
}
\startdata
Haro~2    & 10 32 30 & +54 25 00 & 1994 Nov 1 & C & 193 \\
Haro~2    & 10 32 30 & +54 25 00 & 1995 May 2 & D &  22 \\
Haro~4/26 & 11 05 00 & +29 09 00 & 1994 Nov 1 & C & 202 \\
Haro~4/26 & 11 05 00 & +29 09 00 & 1995 May 2 & D &  32 \\
\enddata

\end{deluxetable}

\clearpage

\begin{deluxetable}{lccccc}
\tablenum{3}
\tablecaption{Parameters of the \hi\ maps made from the combined C+D array data
\label{tab-obshi}} 
\tablewidth{0pt}
\tablehead{
\colhead{} & 
\colhead{Beam size} & 
\colhead{Beam size} & 
\colhead{$\Delta \nu_{\rm chan}$} & 
\colhead{rms} & 
\colhead{rms in units of $T_{\rm b}$} \\
\colhead{Field} & 
\colhead{(${\rm arcsec^2}$)} & 
\colhead{(kpc)} & 
\colhead{(kHz)} & 
\colhead{(${\rm mJy\,beam^{-1}}$)} & 
\colhead{(K)} \\
\colhead{(1)} & 
\colhead{(2)} & 
\colhead{(3)} & 
\colhead{(4)} & 
\colhead{(5)} & 
\colhead{(6)}
}
\startdata
Haro~2    & 14.8\,$\times$\,14.1 & 1.4   & 96.6   & 0.37  &  1.1 \\
Haro~4/26 & 15.5\,$\times$\,14.0 & 0.5   & 96.6   & 0.39/1.95 &  1.1/5.4 \\
\enddata

\end{deluxetable}


\begin{deluxetable}{lcccc}
\tablenum{4}
\tablecaption{OVRO CO(1--0) observing log for Haro\,2 \label{tab-logco}}
\tablewidth{0pt}
\tablehead{
\colhead{} & 
\colhead{Array} & 
\colhead{Longest} &  
\colhead{Passband} & 
\colhead{1150+497} \\
\colhead{Date} & 
\colhead{config} & 
\colhead{baseline} & 
\colhead{calibrator} & 
\colhead{strength (Jy)} \\
\colhead{(1)} & 
\colhead{(2)} & 
\colhead{(3)} & 
\colhead{(4)} & 
\colhead{(5)} 
}
\startdata
1995 Oct 5  & L & 115\,m   & 3C84       & 1.0 \\
1995 Dec 17 & H & 242\,m   & 0528+134   & 0.8 \\
1996 May 11 & L & 115\,m   & 3C273      & 0.9 \\
\enddata

\end{deluxetable}

\clearpage

\begin{deluxetable}{lrrrrcc}
\tablenum{5}
\tablecaption{Measured \ion{H}{1} and 20--cm radio continuum parameters 
\label{tab-hiparm}}
\tablewidth{0pt}
\tablehead{
\colhead{} & 
\colhead{$v_{\rm c}$} & 
\colhead{${\Delta}\,v_{50}$} & 
\colhead{$F_{\rm H\,I}$} & 
\colhead{$S_{\rm 1.4\,GHz}$} & 
\colhead{$M_{\rm H\,I}$} &
\colhead{$N^{\mathit peak}_{\rm H\,I}$} \\
\colhead{ID} & 
\colhead{(${\rm km\,s^{-1}}$)} & 
\colhead{(${\rm km\,s^{-1}}$)} & 
\colhead{(${\rm Jy\,km\,s^{-1}}$)} & 
\colhead{(mJy)} & 
\colhead{($10^{8}$\,\msolar)} &
\colhead{($10^{21}$\,cm$^{-2}$)}\\             
\colhead{(1)} & 
\colhead{(2)} & 
\colhead{(3)} & 
\colhead{(4)} & 
\colhead{(5)} & 
\colhead{(6)} &
\colhead{(7)}
}
\startdata
Haro 2  & 1443 & 111 &  2.6 & 24.6 & 2.3  & 2.0   \\
Haro 4  &  650 &  43 &  1.1 &  3.4 & 0.2  & 2.4   \\
Haro 26 &  712 & 187 & 40   & 23.8 & 8.5  & 2.2   \\
\enddata

\end{deluxetable}


\begin{deluxetable}{lccccccc}
\tablenum{6}
\tablecaption{Inferred global parameters \label{tab-ratios}}
\tablewidth{0pt}
\tablehead{
\colhead{} &
\colhead{} &
\colhead{$M_{\mathrm {H\,I}}/L_B$} &
\colhead{$M_{\rm dyn}$} &
\colhead{$M_{\rm dyn}/L_B$} &
\colhead{} &
\colhead{$M_{\rm H_2}$} & 
\colhead{} \\
\colhead{ID} &
\colhead{$D_{\rm H\,I}/D_{\rm opt}$} &
\colhead{(\msolar/\lsolar)} &
\colhead{($10^9\,M_\odot$)} &
\colhead{(\msolar/\lsolar)} &
\colhead{$M_{\rm H\,I}/M_{\rm dyn}$} &
\colhead{($10^7\,M_\odot$)} & 
\colhead{$M_{\rm H\,I}/M_{\rm H_2}$} \\
\colhead{(1)} & 
\colhead{(2)} & 
\colhead{(3)} & 
\colhead{(4)} & 
\colhead{(5)} & 
\colhead{(6)} & 
\colhead{(7)} &
\colhead{(8)}
}
\startdata
Haro 2  & 1.4 & 0.09 & ~~2.5  &  ~~1.0 & 0.09 &        7.3 &        3.1 \\
Haro 4  & 2.8 & 0.17 & ~~0.5  &  ~~4.8 & 0.03 & $\leq 1.6$ & $\geq 1.2$ \\
Haro 26 & 1.5 & 0.76 & 11.5   &  10.3  & 0.07 &         --- &        --- \\
\enddata

\end{deluxetable}

\clearpage

{\bf Fig.\,1} Mosaic of the channels containing line emission from Haro~2,
each superposed on a DSS $B$--band greyscale image.  Contours are $-$2.5,
2.5, 5.0, and $10\sigma$; negative contours are dashed.  The beam size
indicated in the upper left panel is $14.8\arcsec \times 14.1\arcsec$.

{\bf Fig.\,2} Global \hi\ profile for Haro 2, obtained by integrating the
channel maps over the area of the source after continuum subtraction,
blanking, and correction for primary beam attenuation. For a comparison with
single--dish \hi\ data, we overplot the profile (labeled HSH95) obtained with
the 100--m Effelsberg telescope by \citet{huch95}

{\bf Fig.\,3} H{\sc i} surface density contours for Haro~2, superposed on a
DSS $B$--band greyscale image.  The contours are 2.0 ($5\sigma$), 4.0, 8.1,
12.2, and $16.2 \times 10^{20}\,{\rm cm^{-2}}$. The size of the synthesized
beam is $14.8\arcsec \times 14.1\arcsec$. Three marginally detected \hi\
clouds which seem to coincide with optical counterparts are identified (A, B,
and C; see Sect.\ \ref{ss-h1inh2}). Solid lines indicate the directions along
which the \hi\ major and minor axes were measured.

{\bf Fig.\,4} Intensity--weighted mean velocity field of Haro~2. The optical
center of the galaxy is indicated with a cross; the numbers give heliocentric
velocity in km\,s$^{-1}$. The synthesized beam is indicated at lower left and
measures $14.8\arcsec \times 14.1\arcsec$.

{\bf Fig.\,5} Integrated CO(1--0) intensity in Haro\,2, with offset molecular
gas concentrations S and NW indicated.  Contours are multiples of $0.52\,{\rm
Jy\,beam^{-1}\,km\,s^{-1}}$, for the $3.3\arcsec \times 2.6\arcsec$
synthesized beam shown at lower left.  Pixels with values of less than
$1\sigma$ were blanked before integrating, and the map has not been corrected
for primary beam attenuation, although the fluxes reported in the text have
been. The cross marks the optical position of Haro~2 listed in Table

{\bf Fig.\,6} Overlay of CO(1--0) contours from Figure~5 on a DSS $B$--band
greyscale image.  The cross marks the optical position of Haro~2 listed in
Table \ref{tab-optparm}.

{\bf Fig.\,7} Overlay of \hi\ contours from Figure~3 on a CO(1--0) greyscale
representation (as in Figure~5). The circle at the lower left indicates the
size of the \hi\ beam.

{\bf Fig.\,8} Overlay of CO(1--0) contours from Figure~5 on the {\it ROSAT}
HRI X--ray greyscale from \citet{summ01}.  The X--ray image has units of
${\rm count\,pixel^{-1}}$ above background, and has been smoothed by
convolution with a Gaussian beam with $\sigma = 3\arcsec$.  The cross marks
the optical position of Haro~2 listed in Table \ref{tab-optparm}.

{\bf Fig.\,9} CO(1--0) channel maps for the velocity range integrated to
produce Figures~5 and 7, after smoothing to $31.2\,{\rm km\,s^{-1}}$
resolution.  Contours are multiples of $11\,{\rm mJy\,beam^{-1}}$ ($\sim
2\sigma$); negative contours are dashed.  The cross marks the optical
position of Haro~2 (listed in Table \ref{tab-optparm}); the ellipse in the
top left panel shows the synthesized beam which measures
3.3\prin\,$\times$\,2.6\prin.  These maps have not been corrected for primary
beam attenuation.

{\bf Fig.\,10} Mosaic of the \hi\ line emission channels of Haro~4 overlaid
on a DSS $B$--band image. Contours are -2.5, 2.5, 5.0, 10, and 15\,$\sigma$;
negative contours are dashed.  The beam size indicated in the upper left
panel is 15.5\prin\,$\times$\,14.0\prin.

{\bf Fig.\,11} \ion{H}{1} surface brightness contours for Haro~4, superposed
on a DSS $B$--band greyscale image.  The contours are 2.1 (5\,$\sigma$), 6.2,
12.3, and $18.5 \times 10^{20}\,{\rm cm^{-2}}$.  The beam size is is
15.5\prin\,$\times$\,14.0\prin.

{\bf Fig.\,12} Mosaic of the channels showing \hi\ line emission from
Haro\,26, superposed on a DSS $B$--band greyscale image. Contours are $-$2.5,
2.5, 5.0, and 10\,$\sigma$; negative contours are dashed. The beam size,
indicated in the upper left panel, is 15.5\prin\,$\times$\,14.0\prin.

{\bf Fig.\,13} Global \hi\ profile for Haro~4 and Haro~26, obtained by
integrating the channel maps over the areas of the sources after continuum
subtraction, blanking, and correction for primary beam attenuation.

{\bf Fig.\,14} \ion{H}{1} surface density contours for Haro~26, superposed on
a DSS $B$--band greyscale image.  The contours are 2.1 (5\,$\sigma$), 6.2,
12.3, and $18.5 \times 10^{20}\,{\rm cm}^{-2}$. The beam measures
15.5\prin\,$\times$\,14.0\prin.

{\bf Fig.\,15} A larger scale \hi\ map of the field including Haro\,4 and
Haro\,26 (at upper left and lower right, respectively), overlaid on a DSS
$B$--band greyscale image.  For clarity, only the two lowest contours from
Figures~11 and 14 are shown.

{\bf Fig.\,16} Intensity--weighted mean velocity field of Haro~26.  The
numbers indicate heliocentric velocity in km\,s$^{-1}$.  The beam size is
indicated by the ellipse and measures 15.5\prin\,$\times$\,14.0\prin.

{\bf Fig.\,17} The radio continuum emission for Haro~26, obtained by
averaging the line--free channels of the \hi\ data cube.  Contours are 2.5,
5.0 and $7.5\sigma$ ($1\sigma \sim 0.1\,{\rm mJy\,beam^{-1}}$) and are
superposed on a DSS $B$--band greyscale image.  The beam is indicated in the
lower left corner and measures 16.8\prin\,$\times$\,15.6\prin.

\end{document}